F. Piacentini*, A. Avella, P. Traina, L. Lolli, E. Taralli, E. Monticone, M. Rajteri, D. Fukuda, I. P. Degiovanni, and G. Brida

# Towards joint reconstruction of noise and losses in quantum channels

**Abstract:** The calibration of a quantum channel, i.e. the determination of the transmission losses affecting it, is definitely one of the principal objectives in both the quantum communication and quantum metrology frameworks. Another task of the utmost relevance is the identification, e.g. by extracting its photon number distribution, of the noise potentially present in the channel. Here we present a protocol, based on the response of a photon-number-resolving detector at different quantum efficiencies, able to accomplish both of these tasks at once, providing with a single measurement an estimate of the transmission losses as well as the photon statistics of the noise present in the exploited quantum channel. We show and discuss the experimental results obtained in the practical implementation of such protocol, with different kinds and levels of noise.

*Corresponding Author: F. Piacentini: INRIM - Istituto Nazionale di Ricerca Metrologica, strada delle Cacce 91, 10135 Torino (Italy), E-mail: f.piacentini@inrim.it
A. Avella, P. Traina, L. Lolli, E. Taralli, E. Monticone, M. Rajteri, I. P. Degiovanni, G. Brida: INRIM - Istituto Nazionale di Ricerca Metrologica, strada delle Cacce 91, 10135 Torino (Italy)
D. Fukuda: NMIJ/AIST, National Metrology Institute of Japan, National Institute of Advanced Industrial Science and Technology, Umezono 1-1-1, 305-8563 Tsukuba, Ibaraki (Japan)

# 1 Introduction

Some of nowadays hottest research fields as quantum metrology and sensing [1–7], quantum information [8–11] and foundation of Quantum Mechanics [12–16] find a fundamental tool in quantum channels, transmitting the information carriers (usually single photons or entangled photon pairs).

An ideal quantum channel should be completely transparent, in order to grant the transmission of all the information passing through it, and free from any noise deteriorating the transmission quality. Unfortunately, a real quantum key distribution (QKD) attempt is always affected by losses, be it via open air (earth-to-earth or earth-to-space) [17–19], because of the interaction of the photons with the atmosphere, or through optical fibers [20–23], because of the fact that single photons can be absorbed and noise photons may appear from scattering processes due to the presence of conventional communication in the adjacent channels.

This is why the characterization of a quantum channel [24–28], especially the determination of its transmission losses and of the noise potentially present in it, is definitely a fundamental task for developing and incrementing the performances of the rising quantum technologies [29–32].

Inspired by some theoretical works [33], we present a protocol giving an estimate of the losses in a quantum channel distributing single photons and, within the same measurement, extracting the photon number distribution [34–36] of the noise in it.

We show experimental results demonstrating reasonably good reconstruction, with fidelities ranging from 90.4% to 99.1%, of different kinds (poissonian and thermal) and levels of noise, together with a good estimate of the channel losses. Even though there are techniques obtaining better results in determining the transmission losses or the noise presence in a quantum channel, up to our knowledge this is the only technique able to perform both tasks at the same time, simplifying (after adequate improvement) the channel characterization process.

# 2 Theoretical framework

Let us assume to have a single-photon source (SPS) producing single photons and distributing them in a lossy and potentially noisy quantum channel. Our aim is the characterization of such channel, i.e. the evaluation of both the amount of losses and the photon statistics of the background noise present, allowing, in determinate scenarios, to identify the potential noise source.

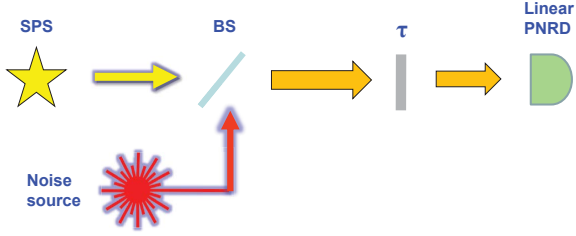

**Fig. 1.** Representation of a noisy and lossy quantum channel connecting a single-photon source (SPS) to a linear photon-number-resolving detector (PNRD): the channel noise is depicted as an unknown photon source mixed with the single photons in a beam splitter (BS), while the losses are simulated by an attenuator of transmittance $\tau$.

From the quantum mechanical perspective, this can be depicted as our SPS output being mixed in a non-polarizing beam splitter with an unknown noise source (see Fig.1) whose photon number distribution is described by the coefficients $\{b(m), m \in \mathbb{N}\}$, where $b(m)$ represents the probability of having $m$ incoming noise photons ($\sum_{m=0}^{\infty} b_m = 1$). After such mixing, the probability of having $m$ photons in the output state will be [37]:

$$p(m) = b(m-1)\xi + b(m)(1-\xi), \quad (1)$$

being $\xi$ the probability of having a single photon in the quantum channel input.

That said, if we suppose to put a linear photon-number-resolving detector (PNRD) with efficiency $\eta$ at the output port of the beam splitter (i.e. at the end of the quantum channel distributing our single photons), we can extract the probability $\Pi(k, \eta, \tau_{ch})$ of having $k$ photocounts:

$$\Pi(k, \eta, \tau_{ch}) = \sum_{m \geq k} \frac{m!}{k!(m-k)!} (\tau_{ch}\eta)^k (1 - \tau_{ch}\eta)^{m-k} p(m). \quad (2)$$

The parameter $\tau_{ch}$, representing the transmissivity of the quantum channel, can be reconstructed, together with the noise photon number distribution coefficients $b(m)$'s, with a single measurement procedure, by exploiting a regularized least-squares minimisation algorithm [33] based on the photocount probabilities $P_i(k)$ registered by the PNR detector for a properly chosen set of different detection efficiencies $\{\eta_i\}$. This result can be achieved, for example, by minimizing the quantity $\sum_{i,k} [P_i(k) - \Pi(k, \eta_i, \tau_{ch})]^2$, with the physical "smoothness" constraint implemented by means of a quadratic, convex and device independent function [38–40].

Being impossible to perform such photon statistics reconstruction in a theoretically infinite Hilbert space, we have to limit ourselves to a truncated reconstruction Fock space, carefully choosing a $M$ for which the probability of having $m > M$ results negligible [35, 36].

## 3 Experimental setup

In our implementation (see Fig. 2), the single photons are produced by a quasi-noiseless heralded SPS [41]. In such setup, a CW laser at 532 nm pumps a $10 \times 1 \times 10$ mm periodically-poled lithium niobate (PPLN) crystal, producing type-0 Parametric Down-Conversion (PDC). The heralding idler photon ($\lambda_i$ = 810 nm) goes through an interference filter (IF, 10 nm FWHM) and an iris, and then is coupled into a single-mode fiber (SMF) connected to a heralding detector, a prototype module [42, 43] based on a red-enhanced silicon-based single-photon avalanche diode [44, 45] (RE SPAD in Fig. 2) designed to provide a high detection efficiency ($\approx 40\%$) and a low timing jitter of $\approx$ 90 ps FWHM at 810 nm. The heralded signal photon ($\lambda_s$ = 1550 nm) is spatially and spectrally filtered by an iris and an IF (30 nm FWHM) respectively, and then it is coupled to a 20 m long SMF connected to a high-speed $2 \times 2$ optical switch (OS) based on a $LiNbO_3$ integrated waveguide Mach-Zehnder interferometer [46], working as an optical shutter. The OS is controlled by a custom-made circuit receiving the heralding signal from the RE SPAD and hence triggering a custom-made fast pulse generator, opening the output channel of the SPS for few nanoseconds in presence of the heralded single photon. This source shows a heralding efficiency of $\xi = (9.2 \pm 0.6)\%$. Furthermore, to investigate the multi-photon component of the SPS we connect it to a Hanbury-Brown and Twiss interferometer, composed of a 50:50 fiber beam splitter (FBS) with the output ports connected to two Indium/Gallium Arsenide single-photon avalanche diodes (InGaAs SPADs), and evaluate the parameter $\alpha = \frac{P_{12}}{P_1 P_2}$ (directly related to the Glauber second-order autocorrelation function $g^{(2)}(0)$) [41, 47, 48], where $P_i$ ($i = 1, 2$) is the photocount probability of the $i$–th InGaAs SPAD and $P_{12}$ is the probability of observing a coincidence count in both the InGaAs SPADs. With this procedure we observe $\alpha = 0.005 \pm 0.007$, among the lowest present in literature.

The output of the SPS is then combined with the noise source in a 99:1 FBS. The noise source is composed of a pulsed laser at 1550 nm triggered by the heralding signal coming from the SPS, addressed to a fiber coupler and then injected in the FBS. A rotating ground glass disk can be inserted before the fiber coupler, in order to change the nature of the induced noise from Poissonian to pseudo-thermal.

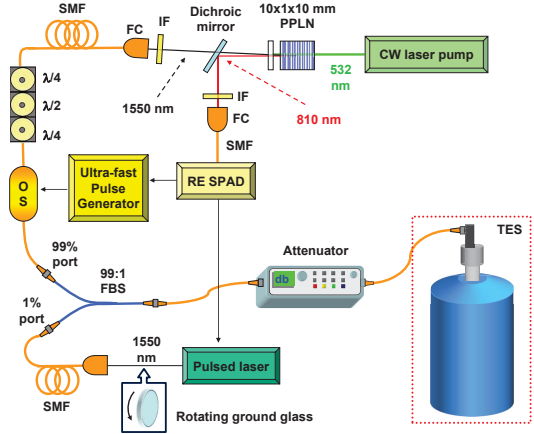

**Fig. 2.** Scheme of the experimental setup. The signal (heralded single photons produced by Type-0 PDC in a PPLN crystal) are mixed with the "noise" photons in a 99:1 FBS. The FBS output is addressed to the detection apparatus, composed of a variable attenuator and a TES-based PNRD.

Being losses critical for the SPS, while not relevant for the noise source, the single photons will enter in the 99% port of the FBS, and the noise photons in the 1% one. We stress that, to avoid unwanted interference effects, the single photons and the noise ones arrive at the FBS with a few nanoseconds time mismatch, still remaining indistinguishable by the adopted detection device.

The output of the FBS goes to a variable attenuator, connected to a superconducting PNRD based on a Transition Edge Sensor (TES) [40, 49, 50]. The TES exploited is composed of a ~ 34 nm thick Ti/Au film [50–52], fabricated by dc-magnetron sputtering. The sensitive area, obtained by lithography and chemical etching, is $10 \times 10$ $\mu$m (more details on the fabrication process are given in [50]). Upon varying the top Ti film thickness, the critical temperatures of these TESs can range between 90 mK and 130 mK, showing a sharp transition (1-2 mK). In order to take advantage of the negative electro-thermal feedback, providing the possibility to obtain a self-regulation of the bias point without a fine temperature control and reducing the detector response time, the TES is voltage biased. In a calibration performed with a PNRD-generalized version of Klyshkho's absolute calibration method [37], the TES detector exploited for the experiment showed a detection efficiency $\eta_{TES} = (67.0 \pm 0.7)\%$. The losses of the last fiber, directly connected to the TES, are included in this value. The attenuator is used to variate the transmittance of the TES input channel, simulating a set of different detection efficiencies: the response of the TES-based PNRD for each efficiency $\eta_i = \eta_{tot}\tau_i$, $(i, 1, ..., N)$ is registered and then used to reconstruct the noise photon number distribution and, at the same time, obtain the estimate of $\eta_{TES}$ by means of a recursive a least-squares minimization algorithm with "smoothness" constraint [35, 36]. The insertion loss of the attenuator was evaluated with classical methods, giving a transmission coefficient $\gamma = 0.76 \pm 0.01$. This means that the overall efficiency of our detection apparatus, formed by the TES and the attenuator, was $\eta_{tot} = \gamma\eta_{TES} = (50.9 \pm 0.9)\%$.

## 4 Obtained results

To test the robustness of our method, we performed six different experimental runs, three with a thermal-like added noise and three with a poissonian one, all with different mean photon numbers per pulse ranging from 0.45 to 2. As predictable, taking into account both the heralding efficiency of the SPS and the global detection efficiency $\eta_{tot}$, our TES was able to discriminate properly only up to $P_i(2)$, with $P_i(3)$ becoming clearly visible only for the acquisitions with the greatest amount of simulated noise (for the evaluation of the experimental photocounts probabilities, see [40]).

Before each acquisition, the whole output of the FBS was connected to a previously calibrated InGaAs SPAD (with efficiency $\eta_{SPAD} = (9.5 \pm 0.2)\%$) able to discriminate between the heralded photons and noise photons contributions, to obtain an estimate of both the channel losses and the added noise mean photon number $\mu$.

In Fig. 3 are reported the plots of the reconstructed photon statistics of the quantum channel noise, compared with the expected ones, for each of the six acquisitions performed. The quality of the reconstructions is generally good, if compared with the expected photon distributions, as certified by the obtained fidelities $F = \sum_{m=0}^{M} \sqrt{b_m^{(r)} b_m^{(e)}}$ (being $b_m^{(r)}$ and $b_m^{(e)}$ respectively the reconstructed and expected $b_m$ elements), 97.8%, 98.3% and 99.1% for the best three cases. For the remaining three ones, as evident from the corresponding plots, the matching between reconstructed and expected photon statistics is not completely satisfactory (as certified by the lower fidelities, ranging from 90% to 96%), demonstrating that our technique is not ready yet for a broad use.

Concerning the quantum channel losses estimation, instead, the average channel transmittance coefficient ex-tracted with our reconstruction method is $\tau_{ch}^{(r)} = 0.84 \pm 0.04$, in excellent agreement with the one measured with the SPAD, $\tau_{ch} = 0.85 \pm 0.02$.

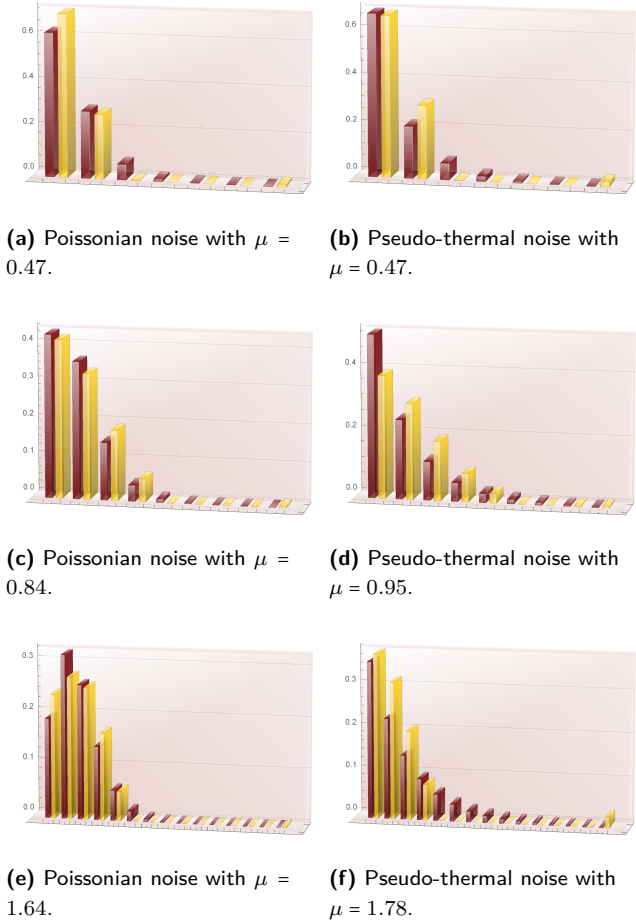

(a) Poissonian noise with $\mu = 0.47$.

(b) Pseudo-thermal noise with $\mu = 0.47$.

(c) Poissonian noise with $\mu = 0.84$.

(d) Pseudo-thermal noise with $\mu = 0.95$.

(e) Poissonian noise with $\mu = 1.64$.

(f) Pseudo-thermal noise with $\mu = 1.78$.

**Fig. 3.** Plots comparing the expected noise photon distributions $b_m^{(e)}$ (brown bars) with the reconstructed ones (yellow bars). The fidelities $F = \sum_{m=0}^{M} \sqrt{b_m^{(r)} b_m^{(e)}}$ obtained are the following: plot (a), $F = 95.7\%$; plot (b), $F = 94.0\%$; plot (c), $F = 99.1\%$; plot (d), $F = 97.8\%$; plot (e), $F = 98.3\%$; plot (f), $F = 90.2\%$.

As a further consistency check, for each acquisition we compared the experimental photocount probabilities $P_i^{(xp)}(k)$ with the ones obtained from the reconstructed photon statistics $(P_i^{(r)}(k))$ and with the expected ones $(P_i^{(e)}(k))$, derived by the $b_m^{(e)}$ coefficients. In Fig. 4 are reported the plots of these quantities (up to $k = 2$) for each of the efficiencies $\{\eta_i\}$, together with the corresponding fidelity between the reconstructed and experimental ones $\left(F_i = \sum_k \sqrt{P_i^{(xp)}(k) P_i^{(r)}(k)}\right)$, in two particolar cases: Poissonian noise with $\mu = 0.84$ in Fig. 4a, pseudo-thermal noise with $\mu = 1.78$ in Fig. 4b. In both cases $F_i$ is always above 99.95%, even if the case of Fig. 4a is the one with the best reconstructed photon statistics ($F = 99.1\%$) while the one represented in Fig. 4b is the one with the least faithful reconstruction outcome ($F = 90.2\%$).

This can be due to the fact that, with such a low overall system efficiency, the $P_i(0)$ contribution is dominant. Hence, the reconstruction method, relying on the contribution of the different photocounts probabilities of the PNR detector to "decouple" the channel transmission losses and the noise photon number distribution coefficients, becomes less reliable, specially when dealing with bigger reconstruction Hilbert spaces (i.e. at higher $\mu$). This means that, in order to achieve a faithful and robust self-characterization of a quantum channel with this method, the whole system efficiency shouldn't drop below few percents, inevitably putting a threshold on the amount of channel losses (and noise level) that one could be able to reconstruct.

In conclusion, we have shown a method for the self-characterization of a quantum channel, able to give an estimate of the transmission losses and of the potential noise photon number distribution at the same time. Even if the results are quite satisfactory and in good agreement with the expectations, we feel that this method still needs to be improved before being considered sound and robust enough for a widespread application.

## Acknowledgments

This work has received funding from EU commission and the EMRP Participating states (EXL02 - SIQUTE project), from H2020 and the EMPIR Participating

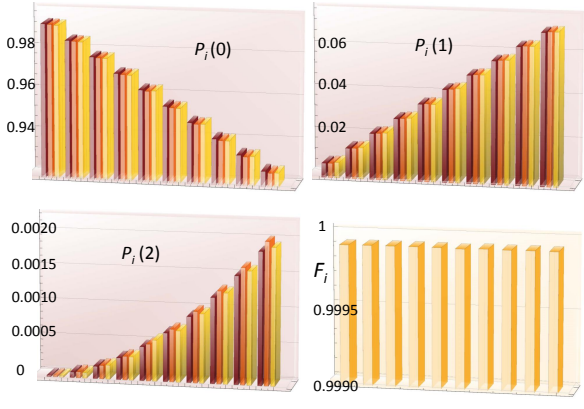

(a) Poissonian noise with $\mu = 0.84$.

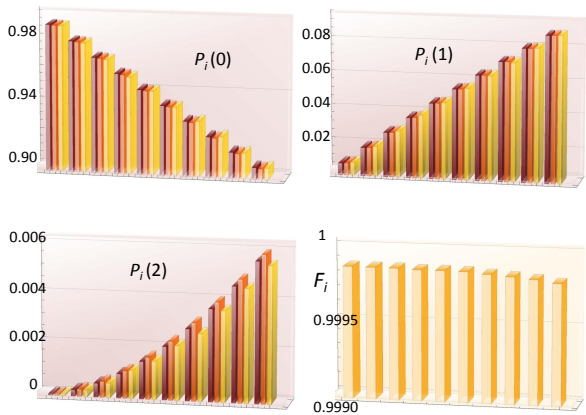

(b) Pseudo-thermal noise with $\mu = 1.78$.

**Fig. 4.** Plots comparing the expected photocount probabilities $P_i^{(e)}(k)$ (brown bars), the reconstructed ones $P_i^{(r)}(k)$ (orange bars) and the ones obtained experimentally $P_i^{(xp)}(k)$ (yellow bars). For each of the two sets, corresponding to two different acquisitions, the bottom-right plot shows the fidelities $F_i$ between $P_i^{(xp)}(k)$ and $P_i^{(r)}(k)$.

states (14IND05 MIQC2 project), and from MIUR (FIRB Project No. D11J11000450001).

We deeply thank Dr. Marco Genovese and Prof. Matteo G. A. Paris for fruitful discussions.